\documentclass[12pt,a4paper]{cibb}

\makeatletter
\providecommand{\@ordinalM}[2]{#1}
\makeatother

\usepackage{subfigure,graphicx}
\usepackage{amsmath,amsfonts,latexsym,amssymb,euscript,xr}
\usepackage{booktabs}
\usepackage[nodayofweek]{datetime}
\usepackage{hyperref}
\usepackage{fmtcount}
\usepackage[english]{datenumber}
\usepackage[absolute]{textpos}
\usepackage{multirow}
\usepackage{graphicx}
\usepackage{multirow}
\usepackage{caption}

\usepackage[table]{xcolor}
\usepackage{color,colortbl,tabularx}

\usepackage[english]{babel}
\usepackage[protrusion=true,expansion=true]{microtype}
\usepackage{amsmath,amsfonts,amsthm}
\usepackage{pifont}

\definecolor{LightBlue}{rgb}{0.88,0.9,0.9}

\title{\Large $\ $\\ \bf A quantum generative model for \textit{in silico} clinical trials using scarce training datasets}

\author{ \large Olatz Sanz Larrarte$^{1,*}$, Reza Dastbasteh$^{1}$,  Roberto Sanchez-Navarro$^{2}$, Maria Diez-Campelo$^{3,4}$, Felipe Prosper$^{4,5,6}$, Ana Alfonso-Pierola$^{4,6}$, Mikel Hernaez$^{2,4,5,7}$, Sara Capponi$^{8,9}$, Pedro Crespo Bofill$^1$, and Josu Etxezarreta Martinez$^1$}
\address{ \footnotesize $\ $\\$^1$ Department of Basic Sciences, Tecnun - University of Navarra, 20018 San Sebastian, Spain. \\
$^2$ CIMA Universidad de Navarra, CCUN, IdisNA, 31008, Pamplona, Spain. \\
$^3$ Hematology Department, Hospital Universitario de Salamanca, Salamanca, Spain.\\
$^4$ Grupo Español de Síndromes Mielodisplásicos (GESMD), Spain. \\
$^5$ CIBERONC, Madrid, Spain. \\
$^6$ Clinica Universidad de Navarra, 31008, Pamplona, Spain. \\
$^7$ DATAI, University of Navarra, 31008, Pamplona, Spain.  \\
$^8$ IBM Research, Silicon Valley, 555 Bailey Ave, San Jose CA 95120 USA. \\
$^9$ Center for Cellular Construction, San Francisco, CA 94158 USA. \\
$^*$Corresponding author: osanzl@unav.es
}

\abstract{\small clinical trials, in silico methods, quantum generative models, data scarcity. \normalsize
\\[17pt]
{\bf Abstract.} In silico methods have emerged as a strategy to complement clinical trials. These are particularly relevant for rare or heterogeneous diseases for which traditional methods are costly or difficult to apply. While classical generative models have shown an extremely good ability to generate high fidelity data when trained using extensive databases, they often struggle when the available samples for training are scarce. In this work, we leverage the potential of quantum computers to represent complex probability distributions to generate high fidelity in silico patients. We propose a pipeline able to combine asymmetric databases into a quantum circuit that serves as a quantum generative model. We evaluate the efficacy of our proposal using a database of Myelodysplastic Syndrome (MDS) patients with 7 clinical variables as a proof-of-concept. We executed our quantum generative model in the IBM Heron r2 ``ibm\_basquecountry'' superconducting quantum computer and compare our method with well known classical baselines. Our results show that the quantum generative model surpasses the classical generative models in generalization and expressivity metrics, indicating its potential validity to generate high fidelity in silico patients for clinical trials.
}

\begin{document}

\renewcommand{\thefootnote}{}
\footnotetext{\small{Article version: \datedate $\;$ h\currenttime  $\;$ CET}}

\thispagestyle{myheadings}
\pagestyle{myheadings}
\markright{\tt Proceedings of CIBB 2026}

\section{Introduction}
\label{sec:SCIENTIFIC-BACKGROUND}


Clinical trials \cite{devereaux2003} remain essential for validating new treatments, but they are often inefficient for small or variable populations. Additionally, traditional cohort studies remain costly and difficult to apply to rare or highly heterogeneous diseases such as myelodysplastic syndromes (MDS) \cite{greenberg2002}. In this context, \textit{in silico} methods have emerged as a complementary strategy \cite{pinero2018silico}. Regulatory agencies are increasingly supporting this transition, as reflected in recent initiatives such as the U.S. Food and Drug Administration (FDA) agency Modernization Act 2.0 \cite{zushin2023fda}.

A critical component in realizing the full potential of \textit{in silico} methodologies is the ability to generate high fidelity synthetic patient data that accurately reflect the underlying clinical variability of real populations. This capability relies on \textit{generative models}, a class of artificial intelligence (AI) algorithms that learn the probability distribution of high-dimensional datasets and can subsequently sample new, realistic observations, usually introducing noise \cite{goodfellow2020generative}.

Despite their promise, \textit{in silico} methods face significant challenges, particularly to ensure model accuracy and reliability. Traditional generative parametric models, e,g. Variational Autoencoders (VAE) or Generative Adversarial Networks (GAN), rely on rigid assumptions that can introduce bias and fail to capture the true complexity of clinical and molecular data distributions. This complexity is further accentuated by the scarcity of available data  (usually involving few hundreds of patients) since classical generative models require large amounts of data for effective training. MDS presents a major clinical challenge due to its complex genomics involving recurrent mutations \cite{rahman2020recurrent} and the lack of consistent effective therapies \cite{kennedy2019genetic}. The high failure rate of Phase III clinical trials in MDS underscores the urgent need for improved patient stratification and trial optimization.

In this context, the advent of quantum computing has led researchers to believe that this novel  paradigm may be successfully applicable for tasks involving data generation \cite{quantumBioRev}. Existing quantum generative models such as Quantum Neural Networks (QNN), quantum VAEs (qVAE) or quantum GANs (qGAN) still show trainability, expressivity or data loading issues that are accentuated in limited training data scenarios \cite{quantumBioRev}. 

In this work, we propose a hybrid quantum-classical pipeline that leverages tensor network representations mapped to quantum circuits to generate high-fidelity synthetic MDS patient cohorts, thereby addressing the dual challenges of data scarcity and high dimensionality inherent in MDS genomics. As a proof-of-concept, we design a quantum generative model for an MDS patient database including $7$ variables which we then execute in the IBM Heron r2 ``ibm\_basquecountry'' quantum processor. To assess the validity of our approach in scarce data scenarios, we use different batch sizes to train the model and evaluate its generalization and expressivity capabilities. We compare the performance of our proposal with three classical generative models: Tabular VAE (TVAE), Conditional Tabular GAN (CTGAN), and CopulaGAN. Our results show the superiority of our pipeline as a generative model for scarce data scenarios.

\section{Data and Methods}
\label{sec:DATA-AND-METHODS}

This study focuses on MDS, a rare and clinically heterogeneous hematological disease for which complete patient data are often limited. The data were obtained from the SintraREV clinical trial, a randomized, double-blind, phase 3 study evaluating low-dose lenalidomide versus placebo in non-transfusion-dependent patients with low-risk del(5q) myelodysplastic syndromes \cite{Sintrarev}. To address the asymmetric availability of clinical variables, the analysis is structured around two complem entary databases: $D_1$, which contains basic clinical and hematological variables routinely collected in practice, namely sex, age, hemoglobin, platelets, and neutrophils, denoted as $\mathbf{x} \in \{0,1\}^5$; and $D_2$, which extends this information by including treatment assignment to lenalidomide and survival outcome, denoted as $\mathbf{y} \in \{0,1\}^2$. This division reflects a common scenario in rare disease research, where comprehensive data collection is challenging and post-treatment variables are often missing or incomplete.


The methodological goal of the project is to generate realistic synthetic patient cohorts that preserve the statistical structure of the observed data while mitigating data scarcity. To this end, we first estimate the target joint distribution $p(\mathbf{x},\mathbf{y})$ of the combined databases $D_1$ and $D_2$ using the maximum entropy principle, which provides the least biased distribution compatible with the empirical information available in the databases. This approach is particularly suitable in rare diseases, where data are incomplete and classical models often require stronger assumptions than the data can support.

Once the target distribution has been estimated, it is transformed into a quantum-compatible representation by introducing an ancilla qubit to encode the probability distribution. The resulting state is then compressed into a Matrix Product State (MPS), which provides a compact and structured representation of the full joint distribution. In this sense, the MPS plays a role analogous to a latent representation in classical generative models, while avoiding strong parametric assumptions about the form of the underlying distribution. The expressive capacity of the MPS is controlled by its bond dimension, which determines how much correlation can be retained and can be adjusted according to the available computational resources. The MPS is then mapped onto a deep quantum circuit following the encoding scheme of \cite{ran2020encoding} and executed on IBM quantum hardware to generate synthetic samples.

Since current quantum devices are noisy and limited in accessible qubit count, we also incorporate a simple error-detection and post-selection step to improve the reliability of the generated samples. Finally, the quality of the synthetic cohorts is evaluated using complementary measures of generalization and expressivity, including distributional similarity metrics and classifier-based tests comparing real and synthetic data. Overall, this pipeline combines classical statistical inference, tensor-network compression, and quantum circuit generation in a hybrid framework designed for synthetic clinical data generation in rare diseases.

\begin{figure}[ht]
    \centering
    \includegraphics[width=0.7\linewidth]{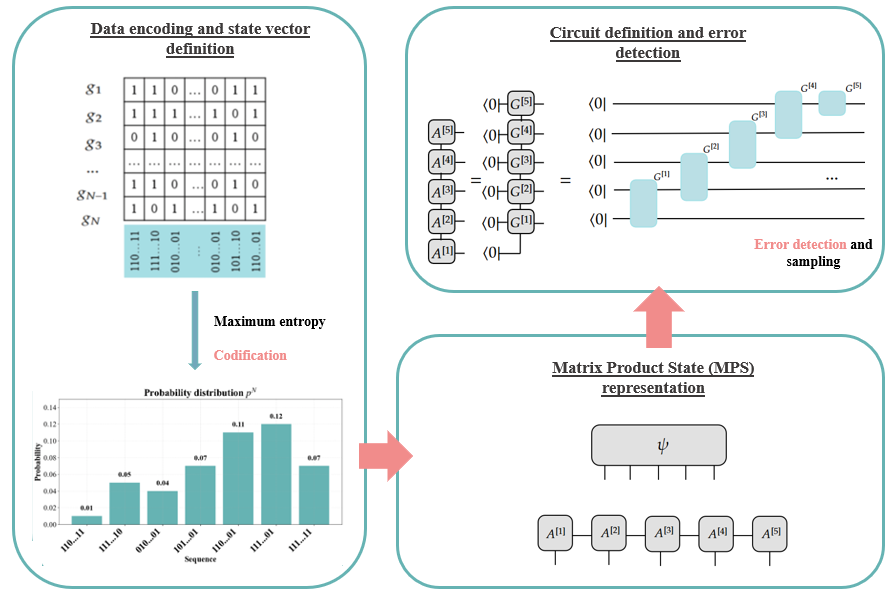}
    \caption{Schematic representation of the methodology for synthetic data generation: (a) data encoding and definition of the joint state vector, (b) MPS discretization of the state, and (c) mapping onto a quantum circuit for sampling.}
    \label{fig:scheme_representation}
\end{figure}

 \subsection{Generalization and expressivity of generative models}

 The quality of generative models is evaluated in terms of two complementary criteria: \textit{generalization} and \textit{expressivity}. Generalization measures how well the model captures the underlying data distribution without memorizing the training set. Expressivity quantifies its ability to generate realistic and diverse synthetic samples. A usual trade-off in generative modeling implies that more expressive models may overfit, indicating poor generalization ability and ``copying'' the training set.

In Table~\ref{tab:metrics}, we summarize the metrics used to assess the quality of the generative models considered in this work. Generalization is quantified using mean absolute error (MAE) and Jensen--Shannon divergence (JSD). Expressivity is evaluated through a random forest test, in which a classifier is trained to distinguish real from synthetic samples on a balanced dataset of 1000 samples (500 real and 500 synthetic) using a 70/30 train--test split. Classification accuracy close to 50\% indicates that the synthetic data are statistically indistinguishable from the real data. In addition, the coverage ratio measures the ability of the generative model to reproduce samples that are present in the target distribution but were not used during training.

 To ensure reproducibility, we report the main hyperparameters used in both the generative models and the evaluation classifier. The Random Forest classifier was configured with 100 trees, Gini impurity, no maximum depth restriction, square-root feature sampling, and default split and leaf settings, together with a fixed random seed in each repetition. The generative baselines TVAE, CTGAN, and CopulaGAN were trained for 400 epochs using the default SDV architecture unless otherwise specified. For the proposed quantum pipeline, the maximum-entropy distribution was encoded using the $[8,7,2]$ classical parity-check code, with a single ancilla qubit and a post-selection step based on syndrome verification.

 To provide a more detailed view of expressivity, we additionally report the corresponding 2$\times$2 confusion matrices of the Random Forest classifier. In this setting, true negatives and false positives quantify how often real samples are correctly recognized as real or incorrectly confused with synthetic ones, whereas false negatives and true positives indicate how often synthetic samples remain indistinguishable from real data or are detected as synthetic.
 


\begin{table}[httb!] \small
\centering
    \begin{tabularx}{\textwidth}{ l  l  c}
        \toprule
          \textbf{Feature} & \textbf{Metric} & \textbf{Interpretation} \\
        \midrule
        \rowcolor{LightBlue} Generalization & Mean Absolute Error (MAE) & Probability distribution difference \\
          Generalization & Jensen-Shannon Divergence (JSD) & Probability distribution difference \\
         \rowcolor{LightBlue} Expressivity & Accuracy & Synthetic sample indistinguishability \\
          Expressivity & Area Under the Curve (AUC) & Synthetic sample indistinguishability \\
         \rowcolor{LightBlue} Expressivity & Coverage Ratio & Model ability to recreate unseen data from target \\
         Expressivity & Confusion Matrix & \shortstack{Error pattern visualization \\ for real-vs-synthetic discrimination}\\
        \bottomrule
    \end{tabularx}
    \caption{\textbf{Metrics used to assess the quality of the generative models}.
    We use multiple metrics to assess both generalization and expressivity of the generative models. We include how to interpret those. \label{tab:metrics}}
\end{table}

Another important aspect of generative models is trainability, i.e. how difficult it is to fit the model before testing. In this work, the model is trained classically using the maximum entropy principle, which allows us to match the empirical distributions extracted from the data. This yields a lightweight training procedure compared with the iterative sample-based optimization commonly required by other generative approaches.

\section{Results}
\label{sec:RESULTS}

\begin{figure}[ht]
    \centering
    \includegraphics[width=0.8\linewidth]{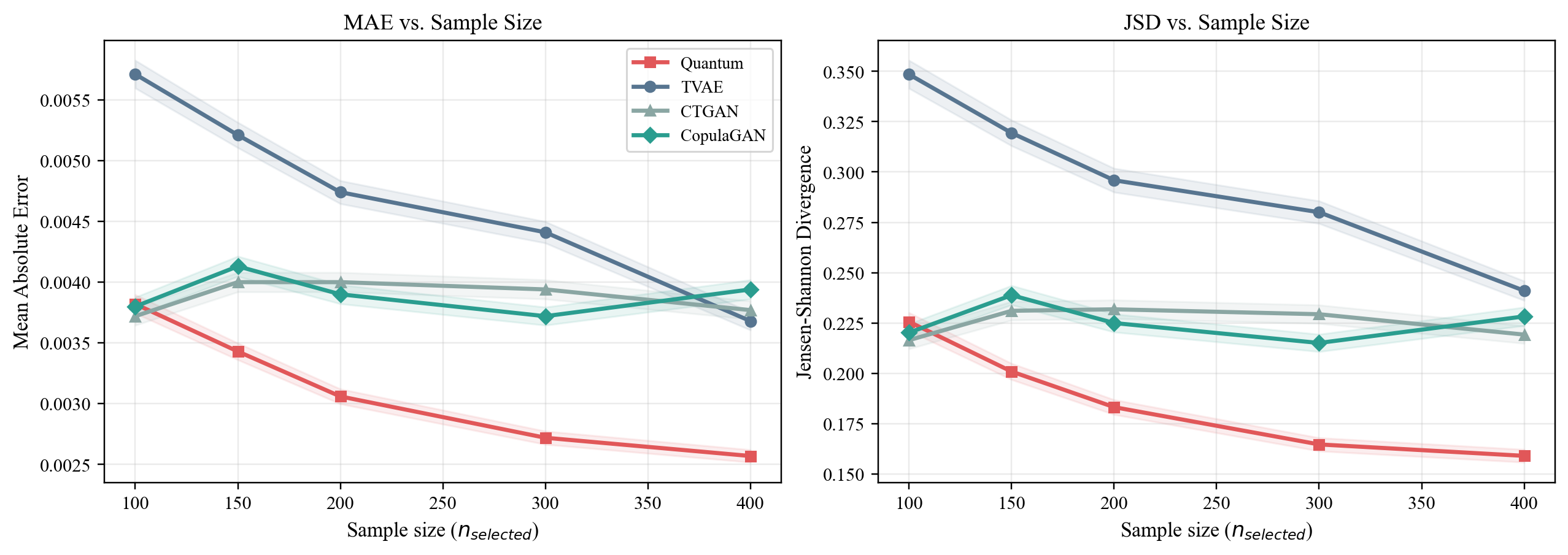}
    \caption{Comparison of the proposed quantum model with classical baselines in terms of MAE and Jensen--Shannon divergence as a function of sample size.}
    \label{fig:mae_jsd}
\end{figure}

We evaluated the proposed quantum synthetic data generation pipeline on the MDS dataset using IBM's ibm\_basquecountry quantum processor, a 156-qubit Heron r2 device. To assess the effect of data availability, we trained the models on progressively larger subsamples, with $n_{\text{\textit{selected}}} \in \{100,150,200,300,400\}$ patients. 

We compare the quantum model with three classical generative baselines: TVAE, CTGAN, and CopulaGAN. As shown in Fig.~\ref{fig:mae_jsd}, the quantum framework achieves better fidelity in terms of MAE and Jensen--Shannon divergence across all sample sizes. In particular, the quantum model consistently improves as the size of the training subset increases, reaching its best performance for the largest batches. While some classical models remain competitive in specific settings, the quantum approach exhibits a more balanced behavior across the full range of sample sizes, suggesting robust generalization under scarce-data regimes.

\begin{figure}[ht]
    \centering
    \includegraphics[width=\linewidth]{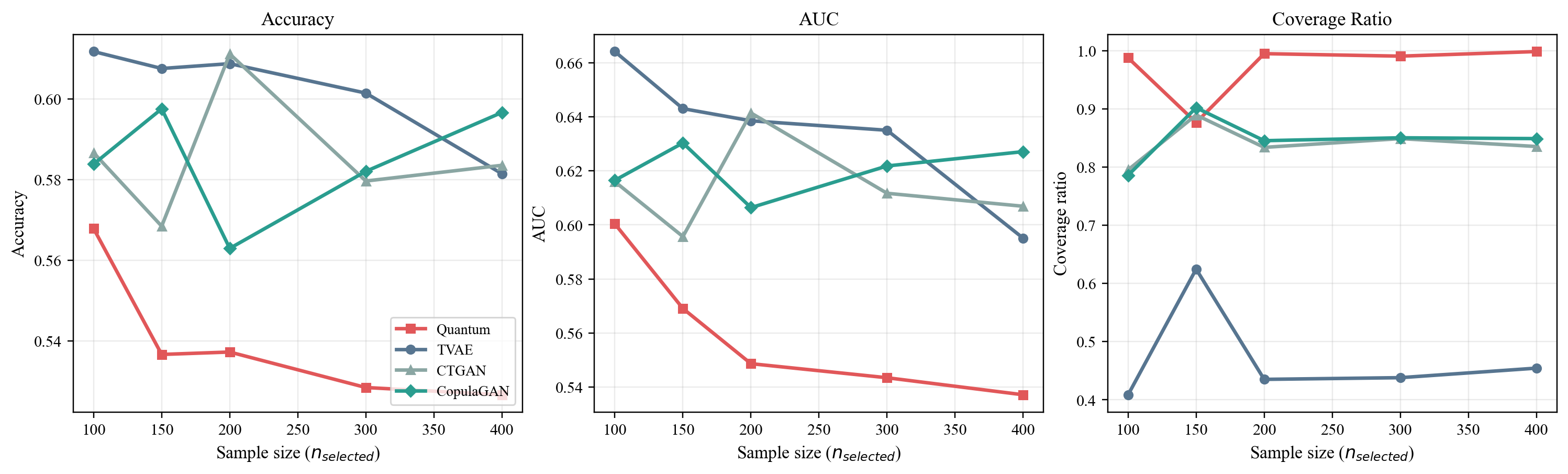}
    \caption{Comparison of the proposed quantum model with classical baselines in terms of classification accuracy, AUC, and coverage ratio as a function of sample size.}
    \label{fig:expressivity}
\end{figure}

To further assess the realism and diversity of the generated samples, we evaluate expressivity through a Random Forest classifier trained to distinguish real from synthetic data, together with the coverage ratio. The results, summarized in Fig.~\ref{fig:expressivity}, indicate that the synthetic data from the quantum model can hardly be distinguished from real data (indicating high quality), while it also achieves the highest or near-highest coverage across all sample sizes (indicating a high expressivity capability). This combination suggests that the model not only captures the global structure of the target distribution, but also reproduces a broad range of distinct patterns, reducing the risk of mode collapse.

\begin{figure}[ht]
    \centering
    \includegraphics[width=\linewidth]{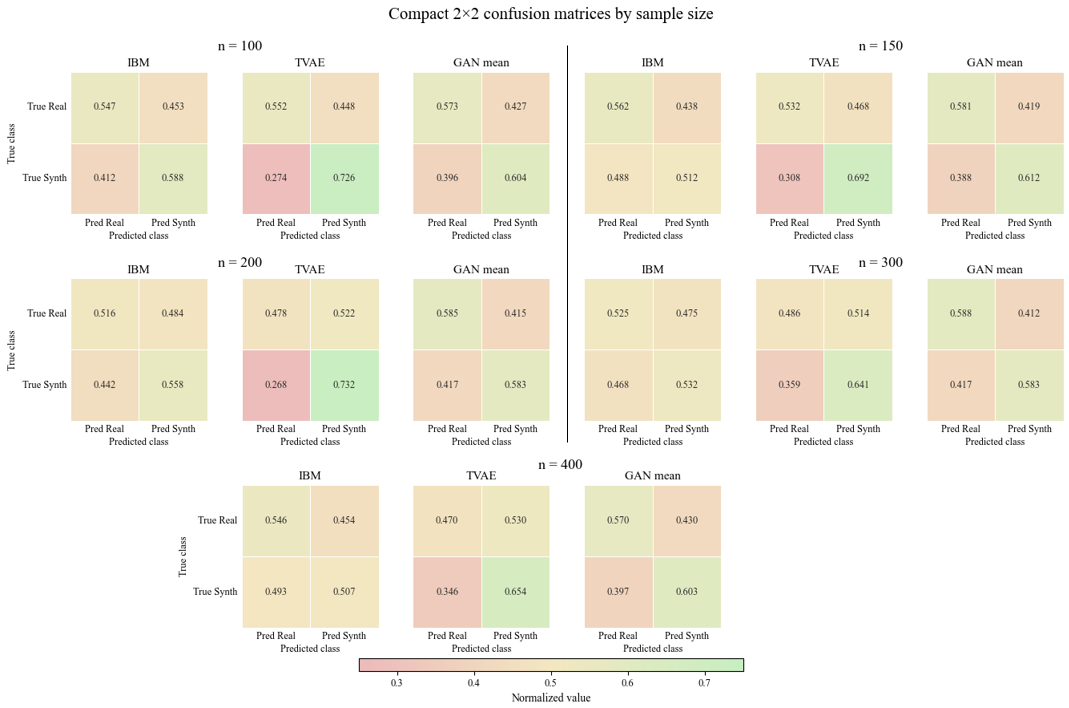}
    \caption{Row-normalized 2$\times$2 confusion matrices of the Random Forest classifier for the proposed quantum model and classical baselines (aggregated GAN baseline obtained by averaging the row-normalized confusion matrices of CTGAN and CopulaGAN) across all sample sizes. Values close to 0.5 indicate near-random discrimination between real and synthetic samples, supporting higher indistinguishability.}
    \label{fig:confusion_matrices}
\end{figure}

To complement these aggregate metrics, Fig.~\ref{fig:confusion_matrices} reports the corresponding 2$\times$2 confusion matrices of the Random Forest classifier for each sample size and generative model. These matrices remain close to chance level for the proposed quantum model, indicating that real and synthetic samples are difficult to distinguish. The classical models show non-trivial biases, indicating the superiority of the proposed quantum generative model.

\section{Conclusion}
\label{sec:CONCLUSIONS}

In this work, we have introduced a quantum generative model specifically tailored to generate high quality \textit{in silico} patients for rare diseases. To do so, we have considered an asymmetric database scenario in which we have access to a huge amount of basic clinical variables, but the amount of prognostically relevant data is extremely scarce. Our pipeline combines both databases using the maximum entropy principle to then find a latent representation of it via MPS tensor networks. The MPS representation is then efficiently translated into a quantum circuit from which high quality synthetic patients can be generated using a quantum computer.

We employed an MDS database to test our proposal using different training batches. This allows us to determine the quality of the generative model for databases consisting of different sizes and compute error bars via randomization. Complementarily, we also trained three classical generative models to determine which of the methods is superior. Our results show that the proposed quantum generative model is the best of the models under examination in terms of generalization and expressivity.

We have also examined the effect of increasing the resolution of the clinical representation by comparing the binarized and discretized encodings (which will be presented in a full paper version of the study). This analysis shows that finer discretization can improve clinical expressivity, but at the expense of larger quantum resources and greater sensitivity to noise. Overall, these results indicate that the proposed hybrid pipeline provides a competitive and flexible framework for synthetic data generation in rare diseases, combining statistical fidelity, sample diversity, and feasibility on current quantum hardware.

\section*{Conflict of interests}
\label{sec:CONFLICT-OF-INTERESTS}
The authors declare no conflict interests.

\section*{Acknowledgments}
\label{sec:ACKNOWLEDGMENTS}
We thank other members of the Quantum Information Lab at Tecnun and the Centro de Investigación Médica Aplicada (CIMA) of University of Navarra  for their support and many useful discussions.

\section*{Funding}
\label{sec:FUNDING}
This work was supported by the Basque Quantum (BasQ) strategy of the Department of Science, Universities, and Innovation of the Basque Government through the “Quantum Synthetic Data Generation for in silico clinical trials (QSynthInSilico)” project.

\section*{Availability of data and software code}
\label{sec:AVAILABILITY}
The data and code that support the findings of this study are available from the corresponding authors upon reasonable request

\footnotesize
\bibliographystyle{unsrt}
\bibliography{bibliography_CIBB_file.bib} 
\normalsize

\end{document}